\documentclass[11pt]{article}

\usepackage[final]{acl}
\usepackage[utf8]{inputenc} 
\usepackage[T1]{fontenc}
\usepackage{amssymb}            
\usepackage{mathtools}          
\usepackage{mathrsfs}           
\mathtoolsset{showonlyrefs}     
\usepackage{graphicx}           
\usepackage{subcaption}         
\usepackage[space]{grffile}     
\usepackage{url}                
\usepackage{xurl}
\usepackage{rotating}
\usepackage{makecell}

\usepackage[most]{tcolorbox}
\usepackage{algorithm}
\usepackage{algpseudocode}
\usepackage{amsmath}
\usepackage{booktabs}
\usepackage{multirow}
\usepackage{threeparttable}
\usepackage{enumitem}
\setlist[enumerate]{leftmargin=*, nosep}
\usepackage{times}
\usepackage{soul}
\usepackage{amsthm}
\usepackage{xcolor}
\usepackage{tabularx}
\usepackage{listings}
\usepackage{textcomp}

\def\email{\small\ttfamily}

\newtcblisting{promptbox}[1][]{
  enhanced,
  breakable,
  colback=gray!5,
  colframe=gray!60,
  title={#1},
  fonttitle=\bfseries,
  listing only,
  listing options={
    basicstyle=\ttfamily\small,
    breaklines=true,
    columns=fullflexible,
    aboveskip=0pt,
    belowskip=0pt
  }
}
\lstdefinelanguage{json}{
  basicstyle=\ttfamily\small,
  breaklines=true,
  frame=single,
  showstringspaces=false
}

\lstdefinelanguage{BetterBash}{
    sensitive=true,
    morecomment=[l]{\#},
    morestring=[b]",
    morestring=[b]',
    breaklines=true,
    frame=single,
}

\title{SW-$A^2$-Bench: Benchmarking Autonomous Software Agent Generation for Agentic Web }

\author{
\normalfont
Linyao Chen$^{2,\dagger}$ \quad
Bo Huang$^{1,4,\dagger}$ \quad
Qinlao Zhao$^{3,\dagger}$ \quad
Shuai Shao$^{1}$ \quad
Zhi Han$^{1}$ \\
Zicai Cui$^{1}$ \quad
Ziheng Zhang$^{5}$ \quad
Guangtao Zeng$^{6}$ \quad
Wenzheng Tang$^{7}$ \quad
Yikun Wang$^{4,8}$ \\
Yuanjian Zhou$^{4}$ \quad
Zimian Peng$^{4,9}$ \quad
Yong Yu$^{1}$ \quad
Weiwen Liu$^{1}$ \quad
Hiroki Kobayashi$^{2}$ \\
Weinan Zhang$^{1,4,*}$ \\[0.5em]
$^{1}$ Shanghai Jiao Tong University \quad
$^{2}$ The University of Tokyo \\
$^{3}$ Huazhong University of Science and Technology \quad
$^{4}$ Shanghai Innovation Institute \\
$^{5}$ Nankai University \quad
$^{6}$ Singapore University of Technology and Design \\
$^{7}$ Queen's University \quad
$^{8}$ Fudan University \quad
$^{9}$ Zhejiang University \\[0.3em]
{\small $^{\dagger}$ These authors contributed equally to this work.
$^{*}$~Corresponding author: \email{wnzhang@sjtu.edu.cn}}
}

\begin{document}

\maketitle

\newcommand{\ourbench}{SW-$A^2$-Bench}
\newcommand{\ourtask}{software agent generation}

\begin{abstract}

The Agentic Web is emerging as a paradigm in which autonomous software agents interact with online resources and with each other to accomplish user goals. However, the capacity of Agentic Web is still limited by insufficient autonomous software agent population, which has become a crucial challenge for scaling Agentic Web. 
In order to alleviate this, we study the task of automatically converting existing code repositories into autonomous software agents via coding agents, decompose the process into critical stages, and identify key technical hurdles. 
To systematically evaluate this capability, we propose \textbf{S}oft\textbf{W}are \textbf{A}gent generation for \textbf{A}gentic Web Bench (\ourbench{}), the first benchmark designed for \ourtask{}. 
\ourbench{} evaluates not only whether software agents can be generated, but also whether generated software agents are faithful to the source repositories and interoperable with other agents in multi-agent workflows.
Our experiments demonstrate that our approach effectively activates the functional capabilities of code repositories and enables interoperable multi-agent collaboration in Agentic Web.
We believe that this work will provide a standardized evaluation for \ourtask{} and will contribute to the future of scaling the capacity of Agentic Web.

\end{abstract}

\section{Introduction}

With the rapid development of Large Language Models (LLMs)~\cite{achiam2023gpt,zhang2024tinyllama}, LLM-based agents~\cite{yao2022react,schick2023toolformer,hu2025owl} have demonstrated remarkable progress in planning~\cite{yao2023treethoughts,wang2023voyager,shen2025satori}, tool use~\cite{schick2023toolformer,liu2025toolace}, and interactive decision-making across a wide range of domains~\cite{deng2023mind2web, wang2025geovista, crab}.
In recent years, LLM-based multi-agent systems (LaMas)~\cite{guo2024large,li2024survey,zhang2025avengers} have attracted increasing attention from the research community, emerging as an effective paradigm for real-world AI applications.
The Agentic Web~\cite{yang2025agentic,guo2025betaweb,zhang2025ufo}, envisioned as the foundational infrastructure for such systems, enables a decentralized ecosystem in which autonomous software agents
interconnect and collaborate to solve complex tasks ~\cite{li2023camel}.
Supported by interoperability standards such as the Agent-to-Agent (A2A) protocol~\cite{a2a_intro} and the Model Context Protocol (MCP)~\cite{mcp-intro}, this direction aims to maximize large-scale collaboration among heterogeneous agents.

\begin{figure}[ht]
    \centering
    \includegraphics[width=0.95\linewidth]{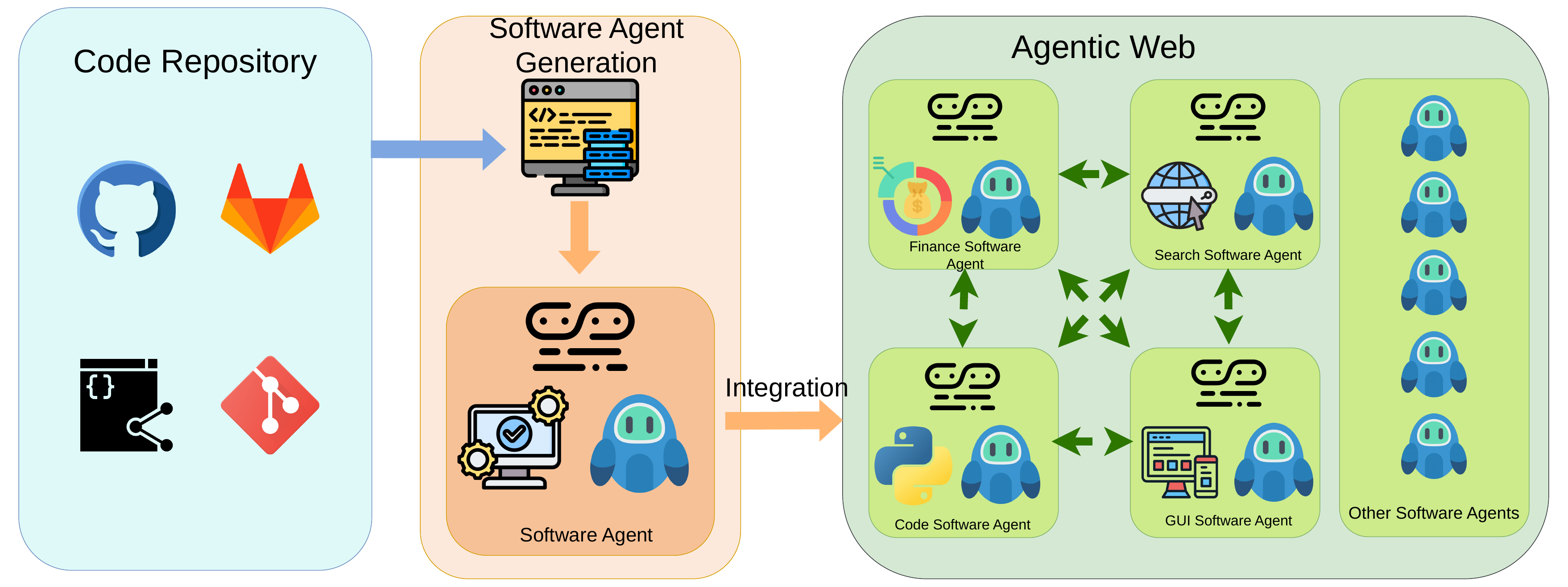}
    \caption{Conceptual illustration of software agent generation for Agentic Web.
    Through agent generation, code repositories are transformed into software agents that can be integrated into the Agentic Web, enabling interactions and collaborations for real-world tasks.}
    \label{fig:teaser}
\end{figure}

As observed in previous work~\cite{kim2025towards,yang2025agentic}, the overall performance of agentic web and multi-agent system (MAS) is largely determined by the available software agent populations. However, current autonomous software agents~\cite{wang2024openhands,yang2024sweagent,wang2025repomaster} are primarily generated through non-standardized manual processes, which requires a large amount of human labor and limits the scaling of software agents and Agentic Web.
To address this, several works~\cite{chen2024autoagents, hu2025automated} have focused on the automated agent generation process, with the aim of increasing the population through non-standardized manual processes.
Considering this, we focus on code repositories, which already encapsulate reusable and domain-specific capabilities, aiming to enrich the software agent population by transforming code repositories.

As illustrated in Figure~\ref{fig:teaser}, in this work, we study the \ourtask{} task which is compliant to communicate in Agentic Web through the A2A protocol, \emph{which is a challenging agentic software engineering task}. 
Compared to ~\cite{jimenez2023swe}, this task is an advanced task which requires higher capacity in software comprehension and reasoning. In addition, the diversity of code repositories in both capacity and programming language makes \ourtask{} a tough task and has hindered previous research~\cite{rashid2025swepolybench,deng2025swe}. What's more, denoted by ~\cite{chen2024internetagentsweavingweb,qian2024chatdev}, to organize an MAS with multiple heterogeneous agents requires adderessing the gaps in organizing heterogeneous architectures, tools, and environments. In our task, with great heterogeneity, it is undoubtedly a challenging agentic task.

To objectively assess progress in this direction, we introduce \ourbench{}, which is, to our knowledge, the first comprehensive benchmark designed to evaluate \ourtask{} for Agentic Web.
Grounded in real-world code repositories, the benchmark targets challenging settings with heterogeneous file formats, complex dependency structures, and the need for autonomous capability verification. \ourbench{} curates 35 diverse repositories with 522 evaluation instances. Specifically, the benchmark evaluates \ourtask{} quality across two critical dimensions: fidelity (the accurate execution of extracted skills) and interoperability (the ability to be reliably discovered and invoked by other agents).
Through this benchmark, we provide a rigorous evaluation of advanced coding agents, identify critical gaps in current approaches, and analyze their strengths and limitations in Section~\ref{subsec:analysis}.

To summarize, our main contributions are as follows:
\begin{itemize}

    \item We systematically formulate automated \ourtask{} process for Agentic Web, clearly define the autonomous software agent format in Agentic Web and frame the process as a repository-level agent task. 

    \item We introduce \ourbench{}, which is, to our knowledge, the first benchmark dedicated to evaluating \ourtask{} for Agentic Web. The benchmark is grounded in real-world software and provides systematic evaluation along fidelity and interoperability dimensions.

    \item We conduct comprehensive experiments by instantiating the \ourtask{} task on representative state-of-the-art coding-agent frameworks. Results indicate that automated \ourtask{} is feasible, while robust environment setup, reliable skill construction, and precise capability specification remain key bottlenecks.
    
\end{itemize}

\section{Related Work}
\label{sec:related_work}

\subsection{Repository-Centric Software Engineering Benchmarks}
Software engineering benchmarks~\cite{jimenez2023swe,badertdinov2025swereben} have increasingly moved beyond isolated programming problems toward repository-centric settings, where models must have a deep understanding and reasoning about complex repositories. One line of work~\cite{liu2023repobench, li2024deveval,li2024evocodebench,li2025feabench} treats repositories as development context for code intelligence, evaluating cross-file retrieval, completion, and repository-level generation. Another line treats repositories as executable workspaces for agentic software engineering, including issue resolution~\cite{jimenez2023swe}, workflow-driven task solving ~\cite{ni2026gittaskbench}, environment setup ~\cite{eliseeva2025envbench}, and long-context codebase reasoning ~\cite{qiu2025locobench}. Recent works~\cite{r2e,aggarwal2026gymanything} further study how software artifacts can be converted into interactive agent environments. Compared with these benchmarks, our work evaluates the construction step that generates standardized software agents with explicit capabilities and interoperable interfaces.

\subsection{Agent Generation Methods}

The software agent generation process endows static contents with autonomous capabilities~\cite{park2023generative, chen2024autoagents}. Early works bridged rigid interfaces with LLM flexibility by transforming services into tools~\cite{qin2023toolllm, liang2024taskmatrix, lin2025autop2c}. Recent efforts directly convert assets into agents, such as transforming scientific papers into interactive MCP servers ~\cite{miao2025paper2agent} or generating executable agents from code repositories ~\cite{wang2025repomaster,chen2025envx}. However, these methods often produce isolated agents, which lack communication capacity and are not compatible with Agentic Web, limiting their broader interoperability.

\subsection{Agent Communication and Coordination}
The Agentic Web~\cite{yang2025agentic} describes a shift from human navigation over linked documents to goal-directed interaction mediated by autonomous agents. This shift is already reflected in user-facing agents for a wide range of tasks~\cite{deng2023mind2web,xie2024travelplanner,huang2025deepresearch}. To support such agents, emerging interfaces and protocols such as MCP\cite{mcp-intro}, and A2A\cite{a2a_intro} expose web resources through natural-language endpoints, tool access, capability discovery, and peer-agent invocation.
Related work~\cite{lu2025buildwebagentsagents,petrova2025semanticweb} further frames the web as an ecosystem of interacting user and content agents. AgentWebBench \cite{zhong2026agentwebbench} instantiates this view by evaluating coordination between a user agent and website-specific content agents. Our benchmark further studies the software agent generation task to expand the software agent population, rather than coordination among existing software agents.

\section{Software Agent Generation Formalization}
\label{sec:method}


In this section, we formulate the task of \ourtask{}.
Given a static code repository ($\mathcal{R}$), the coding agent is tasked with transforming it into an autonomous software agent ($\mathcal{A} = \{\mathcal{A}_{in}, \mathcal{C}\}$) for the Agentic Web ecosystem.

We formalize the context of a code repository as a set of workspace components ($\mathcal{W}$):
\begin{equation}
    \mathcal{R} = \{\mathcal{W}_{dep}, \mathcal{W}_{conf}, \mathcal{W}_{codes}, \mathcal{W}_{readme}\},
\end{equation}
where $\mathcal{W}_{dep}$ and $\mathcal{W}_{conf}$ denote the dependencies and configurations, while $\mathcal{W}_{codes}$ and $\mathcal{W}_{readme}$ represent the core implementation and documentation. This task is decomposed into several subtasks in the following phases:

\textbf{Phase 1: Environment Synthesis.} In this phase, coding agents are required to build corresponding environment $\mathcal{E}$ for original software. The process $\mathcal{P}_{env}$ can be defined as:
\begin{equation}
        \mathcal{E} = \mathcal{P}_{env} (\mathcal{W}_{dep}, \mathcal{W}_{conf})
\end{equation}
which requires coding agents to autonomously resolve implicit system-level conflicts rather than simply executing standard package managers.

\textbf{Phase 2: Semantic Skill Extraction.} In this phase, coding agents are required to extract atomic functional units $\mathcal{S}$ from original software. This subtask $\mathcal{P}_{skill}$ can be defined as follows:
\begin{equation}
        \mathcal{S} = \mathcal{P}_{skill} (\mathcal{R}, \mathcal{E})
\end{equation}
This subtask aims to bridge a profound semantic gap, which requires a thorough comprehension in original software, crucially a challenging task for coding agents.

\textbf{Phase 3: Agent Composition.} In this process, coding agents are required to generate autonomous software agents based on $\mathcal{S}$ and $\mathcal{E}$ generated in previous stage, defined as following:
\begin{align*}
            \mathcal{C} &= \mathcal{P}_{card} (\mathcal{R}, \mathcal{S})\\
        \mathcal{A}_{in} &= \mathcal{P}_{inner} (\mathcal{R}, \mathcal{E}, \mathcal{S})\\
        \mathcal{A} &= \{\mathcal{A}_{in}, \mathcal{C}\}
\end{align*}
where $\mathcal{C}$ denotes the agent card and $\mathcal{A}_{in}$ denotes the inner agent, composing software agents which are available in Agentic Web.

\section{\ourbench{}}
\label{sec:bench}

We introduce the \ourbench{}, a comprehensive dataset and evaluation framework designed to assess the capability of \ourtask{} methods to autonomously transform open-source software repositories into agents that operate within the Agentic Web, interacting with peer agents via the A2A protocol.

\subsection{Benchmark Construction}
The construction of \ourbench{} follows a systematic pipeline. We first curate a diverse set of repositories and establish ground-truth Agent Skills as the static foundation. Subsequently, we generate execution scenarios at two granularities: single-repo and multi-repo tasks. These artifacts collectively underpin our dual-dimensional evaluation: single-repo tasks specifically enable the Capability Inheritance Assessment, while the Agent Skills annotation and multi-repo tasks facilitate the comprehensive Collaborative Execution Assessment.

\paragraph{Repository Collection.}
To rigorously assess agent capabilities across the Agentic Web, we curate 35 diverse GitHub repositories (see Appendix~\ref{sec:repos} for the full list). 
Drawing inspiration from recent repository analysis taxonomies~\cite{ni2026gittaskbench}, we organize these repositories into 9 primary task domains to ensure a broad coverage of real-world software interactions, shown in Figure~\ref{fig:data_distribution}.
This domain-centric selection captures heterogeneous file formats and cross-file transformation workflows, enabling realistic inter-agent dependencies. To support the subsequent \ourtask{} pipeline, these repositories are standardized with rigorous quality control and protocol-compliant transformations.

\paragraph{Agent Skill Annotation.}
Rather than manually synthesizing all metadata fields, we focus strictly on establishing the ground truth for Agent Skills, which are the functional core of generated software agents.
Human experts manually identify and formalize these skills by extracting key functionalities from repository artifacts (e.g., \texttt{README.md}, unit tests).
To ensure these skills faithfully represent the repository's value, we enforce two annotation criteria:
(1) Core Capability Inheritance: The skill must encapsulate a primary, non-trivial feature of the repository (e.g., image segmentation for a vision library) rather than a generic utility;
(2) Atomic Functional Unit: Each skill should represent a distinct, reusable action that can be invoked independently within an agentic workflow.
These expert-annotated skills serve as the gold standard for verifying the specification quality of the generated Agent Card, ensuring each agent uses semantically precise descriptions to be correctly discovered by the orchestrator.
In total, we annotated 127 Agent Skills across 35 repositories.

\paragraph{Execution Task Generation.}
To evaluate \ourtask{} under realistic Agentic Web interactions, we construct execution tasks that are strictly repository-dependent, such that correct solutions must invoke repository-specific skills rather than relying on generic reasoning or external libraries.

We generate tasks using two procedures.
For single-repo tasks (see Appendix~\ref{subsec:single_repo_data_generation_prompt}), a code agent analyzes each individual repository's documentation, unit tests, and exposed APIs to synthesize tasks whose solutions require invoking internal, repository-specific functions, directly probing intra-agent capability inheritance.
For multi-repo tasks (see Appendix~\ref{subsec:multi_repo_data_generation_prompt}), we adopt a two-stage process in which a search agent first identifies semantically complementary repositories, after which a code agent analyzes each repository in isolation and jointly constructs cross-repository workflows.
These workflows are structured as sequential multi-agent chains, where each agent must consume the concrete output of the previous agent as its sole valid input, making every step indispensable. This linear structure disentangles the evaluation of cross-repository interoperability and data-transfer fidelity from higher-order planning complexity.

To ensure task validity and fairness, all generated tasks undergo a verification pipeline.
Each task is executed in a dry run in isolated environments to eliminate execution errors, followed by expert human review to confirm semantic meaningfulness and strict dependence on repository-specific capabilities.
After verification and filtering, we retain 336 single-repo tasks and 186 multi-repo tasks.

\subsection{Execution Task Data Analysis}


\begin{figure*}[htbp]
    \centering
    \includegraphics[width=1\textwidth]{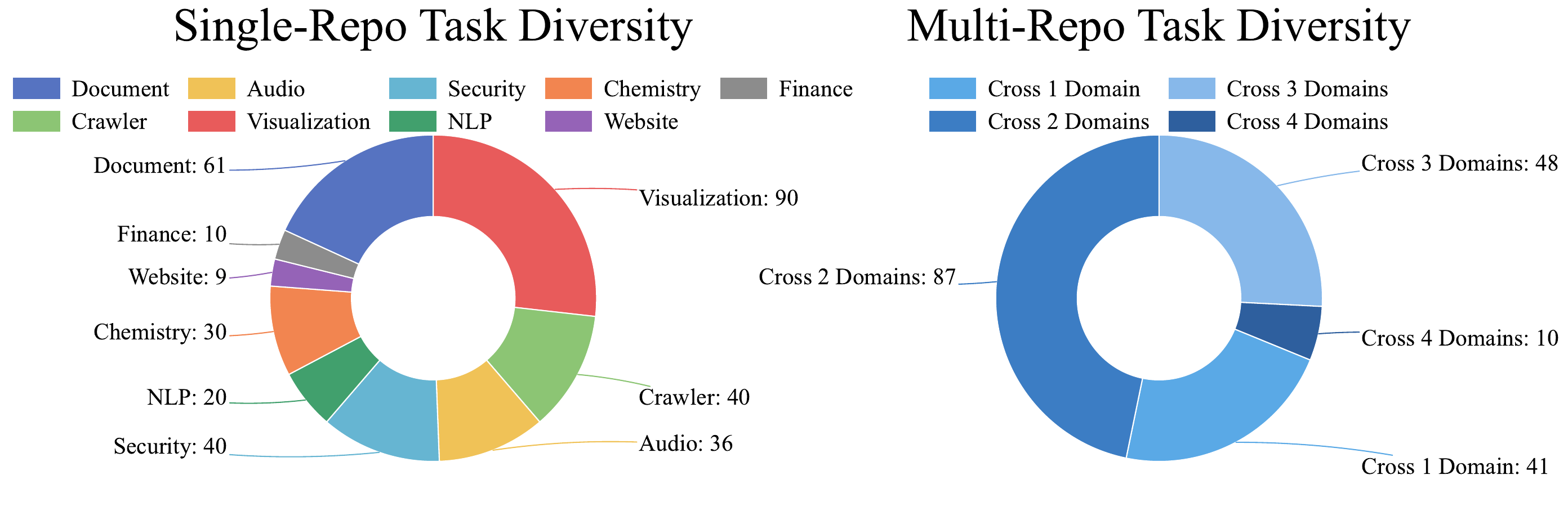}
    \caption{Execution task diversity analysis.
    Single-repo tasks are distributed across 9 application domains, illustrating broad coverage of domain-specific execution scenarios.
    Multi-repo tasks are grouped by the number of distinct domains involved in each workflow (cross-$k$ domains), predominantly highlighting the prevalence of cross-domain interactions and increasing interoperability requirements.}
    \label{fig:data_distribution}
\end{figure*}

\begin{figure}[htbp]
    \centering
    \includegraphics[width=1\linewidth]{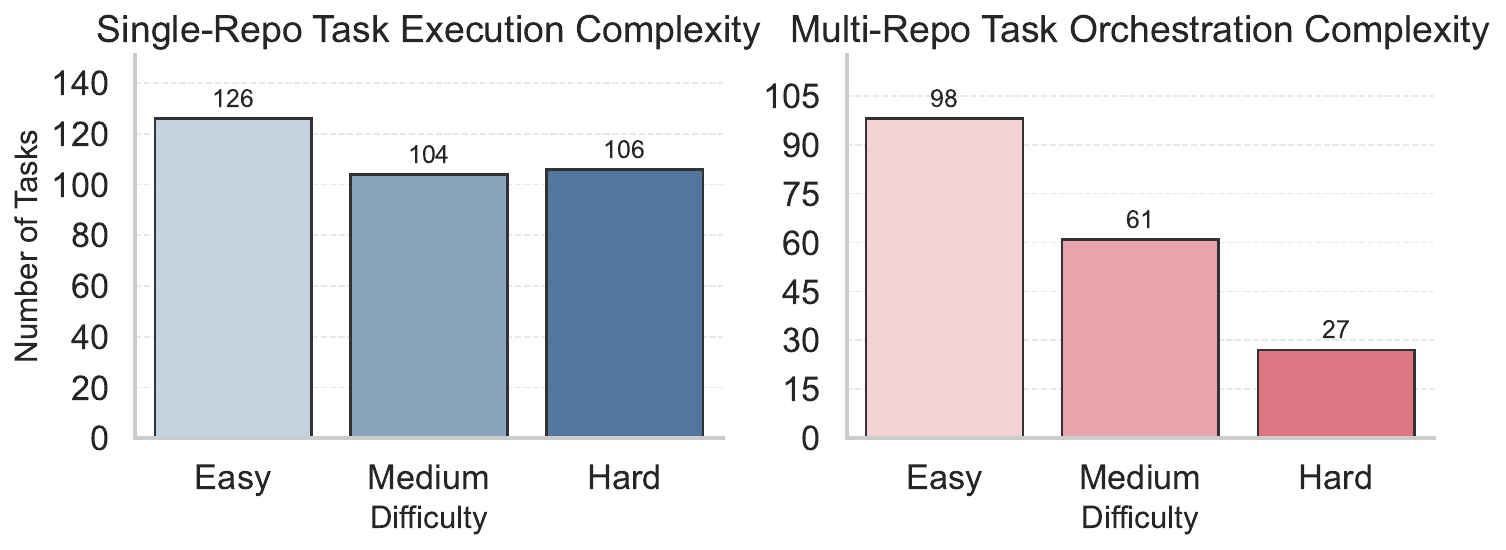}
    \caption{Execution task complexity distribution.
    For single-repo tasks, difficulty is decomposed into several dimensions, measured by corresponding indicators.
    For multi-repo tasks, difficulty is determined by orchestration complexity, measured by the length of the sequential multi-agent chain (i.e., the number of sequentially invoked repositories).
    Tasks are grouped into easy, medium, and hard tiers based on these respective criteria.}
    \label{fig:data_complexity}
\end{figure}

\paragraph{Execution Task Diversity.}
We analyze task diversity along two dimensions: domain coverage in single-repo tasks and cross-domain composition in multi-repo tasks. As shown in Figure~\ref{fig:data_distribution}, the single-repo tasks cover 9 application domains, from vision processing to document parsing, with no category accounting for more than 27\% of the total. Multi-repo tasks test further complex scenarios by requiring agents to orchestrate workflows across 1 to 4 domains, with particular emphasis on the systematic evaluation of cross-domain interoperability.

\paragraph{Execution Tasks Complexity.}
Figure~\ref{fig:data_complexity} summarizes the distribution of task difficulty under two complementary notions of complexity.

To objectively evaluate the execution complexity of single-repo tasks, we design a multi-dimensional indicator system. This framework assesses tasks based on environment setup, output predictability, processing patterns, and domain specificity. We define four binary indicators to capture different facets of task difficulty:
\begin{itemize}
    \item \textbf{$D1$ (Constrained Environment):} Evaluates whether the environment setup requires system-level dependencies, pre-trained models ($>100$MB), or external network access beyond standard package managers.
    \item \textbf{$D2$ (Uncertain Output):} Determines if the task output is non-deterministic due to ML inference variances, external API volatility, or runtime state dependencies.
    \item \textbf{$D3$ (Non-standard Processing):} Identifies tasks that deviate from the standard "Input File $\rightarrow$ Process $\rightarrow$ Output File" transformation pipeline.
    \item \textbf{$D4$ (Domain Expertise):} Measures whether completing the task requires specialized knowledge (e.g., chemistry, finance) beyond general programming and ML engineering.
\end{itemize}
Tasks are categorized into three tiers: Easy (0-1 indicator satisfied), Medium (2 indicators satisfied), and Hard (3+ indicators satisfied).

For multi-repo tasks, we define Orchestration Complexity as the depth of inter-repository coordination required by a workflow.
This measure is derived directly from the sequential multi-agent chain specified during data construction, namely the number of repository-level steps that must be executed in order.
The tasks are grouped into three tiers: Easy (2--3 steps), Medium (4--5 steps), and Hard (6+ steps).

\subsection{Evaluation Pipeline and Metrics}
\label{sec:bench_metric}
To evaluate generated software agents, we perform a comprehensive three-stage assessment covering both the software agent generation process and the quality of generated software agents.

\paragraph{Stage 1: Software Agent Generation Process Assessment.}
We first measure the success rate and cost of transforming original software into autonomous software agents.
\begin{itemize}
\item Software Agent Generation Success (Pass@k): An attempt is considered successful only when the deployed agent's Agent Card is retrievable from the A2A endpoint and passes strict schema validation, including valid, non-empty tool definitions.
\item Software Agent Generation Cost: We record the resource overhead of the \ourtask{} process, specifically reporting Token Consumption per repository.
\end{itemize}

\paragraph{Stage 2: Capability Inheritance Assessment (Single-Agent).} 
This stage validates whether the functional value of the repository is effectively activated and ready for execution.

We assess this through one metric:
\begin{itemize}
    \item Execution Success Rate (SR): We deploy the Software Agent into a sandbox to solve single-repo tasks, which require invoking specific tools defined in the Agent Card.
\end{itemize}

\paragraph{Stage 3: Collaborative Execution Assessment (Multi-Agent).}

This stage evaluates whether an \ourtask{} method can produce a software agent that can be effectively orchestrated under a user-chosen orchestration mechanism in the Agentic Web: specifically, whether the resulting Agent Card makes the agent's unique, task-critical capabilities discoverable so that the mechanism can reliably route the right sub-tasks to it (especially when those sub-tasks can only be solved by that agent) during cross-repository collaboration.

The benchmark is designed to be compatible with orchestration mechanisms in any Agentic Web implementation: it provides a pluggable coordination interface, allowing users to configure and plug in their own orchestration strategy (e.g., platform-native orchestrators, planners, or heuristic policies) as long as it can dispatch sub-tasks to A2A agent endpoints and produce execution traces for evaluation.

We report three key metrics:
\begin{itemize}
    \item Skill Specification Quality: We compare the Agent Card's skill definitions against the annotated ground-truth Agent Skills (see Appendix~\ref{subsec:agent_skill_judge}). We report repository-level Avg-Precision and Avg-Recall as a coarse proxy: Precision reflects whether generated skills are supported by the ground truth, while Recall reflects how completely ground truth skills are covered.
    \item Orchestration Success Rate (Orch. SR): The percentage of multi-repo tasks where all required sub-tasks are correctly dispatched to the appropriate A2A agent endpoints (all-or-nothing, task-level).
    \item Multi-Repo Execution Performance (Exec. SR): The percentage of multi-repo tasks for which orchestration succeeds and the dispatched software agents subsequently complete execution successfully.
\end{itemize}

\paragraph{Metric Implementation.}
All success determinations of execution tasks leverage a standardized LLM-as-a-Judge framework. The judge compares the agent's execution results (including terminal outputs and generated artifacts) against the verified ground truth to determine task success.

\section{Experiments}

This section presents the experimental setup, results and analysis, providing a comprehensive evaluation of the benchmark’s effectiveness, reliability, and practical value in different models and tasks.

\subsection{Experiment Setup}

\paragraph{Coding Agents and Models.}

We evaluate four frontier coding agents that span general-purpose coding, open-source development, and specialized environment configuration.

Claude Code, instantiated with \texttt{Claude-Sonnet-4-5-20250929}~\cite{anthropic2025sonnet45systemcard}, represents the frontier of general-purpose agentic coding.
Codex CLI, driven by \texttt{GPT-5.2-Codex}~\cite{openai2025gpt52codex}, provides a strong open-source baseline.
OpenHands~\cite{wang2024openhands}, a platform for the development of AI agents, configured with \texttt{Claude-Sonnet-4-5-20250929}.
EnvX~\cite{chen2025envx}, also using \texttt{Claude-Sonnet-4-5-20250929}, specializes in autonomous environment configuration and dependency resolution, directly addressing the challenge of environment synthesis which is central to \ourtask{}.

The corresponding task prompts used to instantiate these agents are provided in Appendix~\ref{subsec:agentization_prompt}.

\paragraph{Architecture Configuration.}

\begin{figure*}[htbp]
\centering
\includegraphics[width=0.95\textwidth]{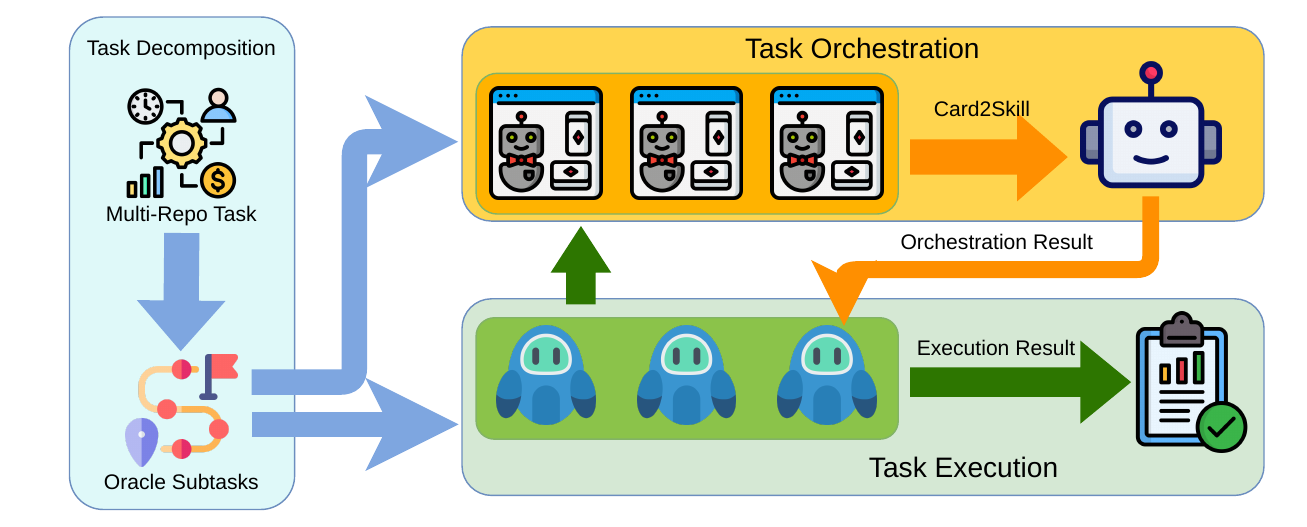}
\caption{
The orchestration mechanism instantiated in our experiments for \ourbench{} Stage 3. 
Multi-repo tasks are provided together with their oracle subtask decompositions obtained during benchmark construction, which serve as the input to the orchestration process. 
In parallel, Agent Cards produced by \ourtask{} agent are converted into agent skills. The coordinator relies solely on these skills to select and bind appropriate software agents for each subtask, orchestrates their execution and obtains the result.}
\label{fig:stage3_orch_trace_eval}
\end{figure*}

To rigorously assess different \ourtask{} methods in a controlled setting, we standardize the agent stack and instantiate a concrete orchestration mechanism (Figure~\ref{fig:stage3_orch_trace_eval}) while keeping the benchmark itself orchestration-agnostic.

\begin{itemize}
    \item Inner Agent Instantiation: Following the definition in Section~\ref{sec:method}, we select Claude Code as the agent scaffold for the Inner Agent ($\mathcal{A}_{in}$). This choice leverages its state-of-the-art tool-use capabilities to effectively drive the extracted repository skills.

    \item Orchestration Instantiation (Claude Code + Oracle Decomposition): For multi-agent evaluation, we instantiate a centralized orchestrator driven by Claude Code (Figure~\ref{fig:stage3_orch_trace_eval}).
        Each multi-repo task is initialized with its oracle subtask decomposition during benchmark construction, and the generated Agent Cards of all A2A agents are converted into executable Claude Agent Skills~\cite{anthropic_agent_skills_overview}.
    The orchestrator then relies solely on these Agent Card-derived skill specifications to route subtasks to the appropriate A2A agents.

\end{itemize}

\subsection{Main Results}
We present our experimental findings following the three-stage evaluation pipeline defined in Section~\ref{sec:bench_metric}.

\paragraph{Stage 1: Agent Generation Process Assessment.} As shown in Table 1, Claude Code and EnvX reliably generate valid agents (100\% Pass@1). Although Codex uses at least 28\% fewer tokens than other baselines, it requires the highest average attempts to generate valid agents, indicating a clear trade-off between token efficiency and agent generation reliability.
\begin{table}[ht]
    \centering
    \caption{Stage 1 Results: Agent Generation Process Assessment. We report Pass@1 and Pass@3 in percentage (\%).}
    \label{tab:agentization_results}
    \small
    {\setlength{\tabcolsep}{1.2mm}%
    \begin{tabular}{lccccc}
        \toprule
        \textbf{Framework} 
        & \textbf{Pass@1} $\uparrow$ 
        & \textbf{Pass@3} $\uparrow$ 
        & \textbf{Tries} $\downarrow$ 
        & \textbf{Tokens} $\downarrow$ \\
        \midrule
        Claude Code & \textbf{100.00} & 100.00 & \textbf{1.000} & 3374449 \\
        Codex & 94.28 & 100.00 & 1.086 & \textbf{2321397} \\
        OpenHands & 94.28 & 100.00 & 1.057 & 3254978 \\
        EnvX & \textbf{100.00} & 100.00 & \textbf{1.000} & 4215051 \\
        \bottomrule
    \end{tabular}}
\end{table}

\paragraph{Stage 2: Capability Inheritance Assessment.} Table 2 demonstrates that Claude Code has the highest overall performance in successfully activating software functions, especially in easy scenarios. Notably, Codex achieves the best success rate in both medium and hard levels, demonstrating Codex's great potential for extracting software functions under more challenging repositories.
\begin{table}[ht]
    \centering
    \caption{Stage 2 Results: Capability Inheritance. We report Execution Success Rate (SR, \%) for task execution.}
    \label{tab:single_repo_results}
    \small
    \begin{tabular}{lcccc}
        \toprule
        \textbf{Framework}
        & \multicolumn{4}{c}{\textbf{Execution SR (\%) $\uparrow$}} \\
        \cmidrule(lr){2-5}
        & \textbf{Easy} & \textbf{Medium} & \textbf{Hard} & \textbf{Overall} \\
        \midrule
        Claude Code & \textbf{57.9} & 38.5 & 10.4 & \textbf{36.9} \\
        Codex       & 45.2           & \textbf{42.3}           & \textbf{14.2}           & 34.5 \\
        OpenHands   & 54.8           & 32.7           & 10.4           & 33.9 \\
        EnvX        & 53.2           & 38.5           & 10.4           & 35.1 \\
        \bottomrule
    \end{tabular}
\end{table}

\paragraph{Stage 3: Collaborative Execution Assessment.} 

\begin{table}[ht]
    \centering
    \caption{Skill Specification Quality: Avg-Precision and Avg-Recall (in \%).}
    \label{tab:skill_metrics}
    \small
    \begin{tabular}{lccc}
        \toprule
        \textbf{Framework} 
        & \textbf{Avg-Precision} $\uparrow$ 
        & \textbf{Avg-Recall} $\uparrow$ \\
        \midrule
        Claude Code & \textbf{70.4} & 70.4  \\
        Codex & 50.0 & 47.8  \\
        OpenHands & 65.2 & 69.7  \\
        EnvX & 69.8 & \textbf{74.0} \\
        \bottomrule
    \end{tabular}
\end{table}

\begin{table}[t]
  \centering
  \caption{Stage 3 Results: Collaborative Execution. We report Orchestration SR (Orch., \%) and Execution SR
(Exec., \%).}
  \label{tab:multi_repo_results}
  \small
  \setlength{\tabcolsep}{5pt}
  \renewcommand{\arraystretch}{0.95}
  \textbf{(a) Orchestration Success Rate (Orch. SR, \%)} \\[4pt]
  \begin{tabular}{lcccc}
      \toprule
      \textbf{Framework}
      & \textbf{Easy} & \textbf{Medium} & \textbf{Hard} & \textbf{Overall} \\
      \midrule
      Claude Code & 81.6 & \textbf{67.2} & \textbf{44.4} & \textbf{71.5} \\
      Codex       & \textbf{83.7} & 59.0 & 25.9 & 67.2 \\
      OpenHands   & 81.6 & 54.1 & 29.6 & 65.1 \\
      EnvX        & \textbf{83.7} & 57.4 & \textbf{44.4} & 69.4 \\
      \bottomrule
  \end{tabular}
  \\[10pt]
  \textbf{(b) Execution Success Rate (Exec. SR, \%)} \\[4pt]
  \begin{tabular}{lcccc}
      \toprule
      \textbf{Framework}
      & \textbf{Easy} & \textbf{Medium} & \textbf{Hard} & \textbf{Overall} \\
      \midrule
      Claude Code & 48.0 & \textbf{39.3} & 7.4 & 39.2 \\
      Codex       & 54.1 & 32.8 & 7.4 & 40.3 \\
      OpenHands   & \textbf{63.3} & 34.4 & \textbf{11.1} & \textbf{46.2} \\
      EnvX        & 61.2 & 23.0 & \textbf{11.1} & 41.4 \\
      \bottomrule
  \end{tabular}
\end{table}

Tables 3 and 4 reveal a clear positive correlation between an agent's specification quality (Skill Avg-Precision and Avg-Recall) and its Orchestration Success Rate.
In complex multi-agent workflows (Hard scenarios), agents with higher-quality skill specifications, such as EnvX and Claude Code, tend to be dispatched more robustly.
Interestingly, OpenHands presents a paradox: it has the lowest Orchestration SR but the highest final Execution SR.
This highlights a critical takeaway: while accurate Agent Cards are essential for the orchestrator to route tasks correctly, a highly robust execution engine (like OpenHands) remains the ultimate safety net for completing complex tasks once dispatched.

\subsection{Analysis}
\label{subsec:analysis}
In this section, we further analyze the performance of software agents, focusing on three failure patterns in software agent generation process.

\subsubsection{Environment Synthesis Analysis}

As indicated by previous work~\cite{eliseeva2025envbench, aggarwal2026gymanything}, environment synthesis can be a challenging problem for current coding agents due to the difficulty in resolving complex system dependencies. In our experiment, the failure of environment synthesis manifests as Startup Failures, A2A Schema Non-compliance, or most frequently, Unusable Skills.
These failures force agents into costly runtime troubleshooting or raw terminal usage (hand-crafting), which significantly inflates token consumption.

This operational overhead is starkly illustrated by the \texttt{aizynthfinder} trajectory, wherein an unconfigured environment forced the agent to expend 6.5 minutes executing 22 distinct package management operations (see Appendix \ref{subsec:case1}).
Such toolchain misalignments ultimately necessitated a disproportionate synchronization of more than 150 packages merely to resolve a single dependency, underscoring the severity of the debugging loop.

\subsubsection{Skill Construction Analysis}
As a key part of our task, skill construction requires both software comprehension and code generation, which suffers from the unverified alignment between generated code and repository implementation. In our experiments, the failures typically appear as hallucinated APIs (non-existent functions) or incorrect signatures.
The \texttt{segment-anything} trajectory exemplifies this fragility: the \texttt{segment\_with\_points} tool repeatedly rejected valid semantic inputs due to a type signature mismatch (expecting objects but receiving strings) (see Appendix \ref{subsec:case2}). 
Unable to resolve this misalignment, agents are forced to abandon encapsulated skills in favor of unstable raw code generation.
As demonstrated in this trajectory, all calls to the segmentation tool failed due to similar implementation defects. Consequently, agents completely bypassed the abstractions originally designed to improve efficiency.

\subsubsection{Capability Specification Analysis}
Agent Cards play an important role in defining communication within the Agentic Web, serving as the functional profiles of the software agent for peer agents.
Our experiments reveal the failure patterns of Agent Cards and how these failures create bottlenecks for communication and collaboration in the Agentic Web.
The issue is particularly evident in OpenHands (Table~\ref{tab:multi_repo_results}), where dispatch failures arise from two distinct mechanisms.
First, semantic blurring led to the incorrect deployment of \texttt{speechbrain\_agent}. Its overly broad capability description overshadowed the more specialized \texttt{spleeter\_agent} (see Appendix~\ref{subsec:case3}).
Second, structural indistinguishability arose when \texttt{chemlib} and \texttt{chemformula} exposed identical ``Generic software agent'' profiles, making them difficult to distinguish based on their Agent Cards alone.
These ambiguities force the orchestrator to abandon deterministic role matching in favor of error-prone heuristic guessing, thereby limiting the robustness of the Agentic Web.

\section{Conclusion and Future Work}

In this paper, we study the process of autonomous software agent generation for scaling Agentic Web.
We outline a basic \ourtask{} workflow, propose the first benchmark for this task, and use it to systematically examine the difficulties encountered when activating real-world software.
Our extensive experiments reveal that current advanced coding agents still struggled in \ourtask{}, limited by several challenges. For future research, this study provides a roadmap to bridge the gap between isolated softwares and a fully interconnected Agentic Web.

\section{Limitations}

This work has several limitations.
First, although our benchmark covers 35 diverse GitHub repositories across multiple domains, it still represents only a limited slice of the broader software ecosystem, and may not fully capture closed-source projects, enterprise-scale systems, highly interactive services, or repositories with complex deployment constraints.
Second, our experiments instantiate collaborative evaluation with a specific A2A-based communication substrate and a centralized orchestration setup, so the observed results may vary under other protocols, decentralized coordinators, or different task-decomposition strategies.
Finally, the evaluated coding-agent frameworks and model backbones are rapidly evolving, so our results should be interpreted as a snapshot of current repository agent generation capabilities rather than a fixed ranking of systems, which may be a potential risk for evaluation.
\bibliography{reference}

\clearpage
\appendix

\section{Case Study}
\label{sec:case_study}

In this section, we present a collection of representative execution trajectories and examples that illustrate the diverse failure modes and decision-making challenges analyzed in the case studies.

\subsection{Case 1: Environment Pre-configuration}
\label{subsec:case1}
\begin{lstlisting}[language=bash,caption={Aizynthfinder Agent Environment Configuration Trajectory (Simplified)},label={lst:case1},basicstyle=\ttfamily\tiny,breaklines=true,frame=single]
# ------------------------------
# 1. Attempting Standard Pip (Failed)
# ------------------------------
# Error: No such file or directory
[VENV_PATH]/bin/pip install python-docx

# Error: No such file or directory
.venv/bin/pip install python-docx

# Error: No module named 'pip' inside venv
.venv/bin/python3 -m pip install python-docx

# ------------------------------
# 2. Switching to 'uv' Tool & Partial Fixes
# ------------------------------
# Verified 'uv' package manager availability
which uv

# Installed successfully, but environment mismatch occurred later
uv pip install python-docx

# ------------------------------
# 3. Manual Path Patching
# ------------------------------
# Action: Manually injected library path into run_retroplanning.py
# Code inserted: sys.path.insert(0, "/path/to/aizynthfinder")

# Failed: Fixed first import error, but hit ModuleNotFoundError for 'rdchiral'
python3 run_retroplanning.py

# Cleanup: Removed lines 17-19 to revert to a clean state
sed -i '17,19d' run_retroplanning.py

# ------------------------------
# 4. Dependency Reset
# ------------------------------
# Attempted to install current package (failed/incomplete)
source .venv/bin/activate && uv pip install -e .

# Major Action: Triggered massive uninstall (170 pkgs) and reinstall (15 pkgs)
uv sync

# Re-installed core dependencies after sync
source .venv/bin/activate && uv pip install rdchiral rdkit aizynthfinder

# Re-installed 'python-docx' into the specific venv python
uv pip install python-docx --python .venv/bin/python3
\end{lstlisting}

\subsection{Case 2: Skill Construction}
\label{subsec:case2}

\begin{lstlisting}[language=json,caption={The provided trajectory exemplifies the fragility of skill construction.},label={lst:case2},basicstyle=\ttfamily\tiny,breaklines=true]
{
    "index": 1,
    "timestamp": "2026-01-07T15:06:00.574575",
    "type": "AssistantMessage",
    "data": {
    "content": [
        {
        "text": "I'll segment the dog from the background using the Segment Anything Model with the point prompt you provided."
        }
    ],
    "model": "claude-sonnet-4-5-20250929",
    "parent_tool_use_id": null,
    "error": null
    }
},
{
    "index": 2,
    "timestamp": "2026-01-07T15:06:01.596017",
    "type": "AssistantMessage",
    "data": {
    "content": [
        {
        "id": "toolu_011zcKdQ6vnc2gRjJhREtDpi",
        "name": "mcp__repo__segment_with_points",
        "input": {
            "image_path": "/tmp/a2a_agent_05gvew92/dog.jpg",
            "point_coords": "[[500, 400]]",
            "point_labels": "[1]",
            "model_type": "vit_h"
        }
        }
    ],
    "model": "claude-sonnet-4-5-20250929",
    "parent_tool_use_id": null,
    "error": null
    }
}
\end{lstlisting}

\subsection{Case 3: Capability Specification}
\label{subsec:case3}

\begin{lstlisting}[language=json,caption={Spleeter Agent Agentcard configured by OpenHands},label={lst:case3a},basicstyle=\ttfamily\tiny,breaklines=true]
{
  "capabilities": {
    "streaming": false
  },
  "defaultInputModes": [
    "text/plain",
    "audio/mpeg",
    "audio/wav"
  ],
  "description": "AI-powered audio source separation agent using Spleeter.",
  "name": "spleeter_audio_separation_agent",
  "version": "1.0.0"
}
\end{lstlisting}
\section{Repositories Used to Build \ourbench{}}
\label{sec:repos}

The repositories are organized by domain, as described below.

\subsection*{1) Document and Web Parsing with OCR}

\begin{itemize}
  \item \textbf{Tesseract}: A classic open-source OCR engine primarily written in C++, featuring multilingual support and a large ecosystem.\\
  \quad GitHub: \url{https://github.com/tesseract-ocr/tesseract}

  \item \textbf{EasyOCR}: An out-of-the-box OCR solution based on PyTorch, supporting more than 80 languages and suitable for rapid integration.\\
  \quad GitHub: \url{https://github.com/JaidedAI/EasyOCR}

  \item \textbf{textract}: A unified interface for extracting plain text from almost any document format (e.g., Word, PDF), often relying on external engines.\\
  \quad GitHub: \url{https://github.com/deanmalmgren/textract}

  \item \textbf{tika-python}: Python bindings for Apache Tika, enabling text and metadata extraction via the Tika REST service.\\
  \quad GitHub: \url{https://github.com/chrismattmann/tika-python}

  \item \textbf{camelot}: A library for structured table extraction from PDF documents.\\
  \quad GitHub: \url{https://github.com/camelot-dev/camelot}

  \item \textbf{python-docx2txt}: A utility for extracting text and images from \texttt{.docx} files.\\
  \quad GitHub: \url{https://github.com/ankushshah89/python-docx2txt}

  \item \textbf{Unstructured}: An LLM-oriented document processing pipeline supporting ETL, chunking, and structured representations.\\
  \quad GitHub: \url{https://github.com/Unstructured-IO/unstructured}
\end{itemize}

\subsection*{2) Web and Platform Scraping for Text Acquisition}

\begin{itemize}
  \item \textbf{trafilatura}: A tool for extracting main content and metadata from web pages, supporting command-line and batch processing.\\
  \quad GitHub: \url{https://github.com/adbar/trafilatura}

  \item \textbf{yt-dlp}: A multi-site audio and video downloader supporting YouTube and other platforms.\\
  \quad GitHub: \url{https://github.com/yt-dlp/yt-dlp}

  \item \textbf{youtube-transcript-api}: A library for retrieving YouTube subtitles and auto-generated captions without requiring an API key.\\
  \quad GitHub: \url{https://github.com/jdepoix/youtube-transcript-api}

  \item \textbf{MediaCrawler}: A crawler framework for collecting content and comments from platforms such as Xiaohongshu, Douyin, Kuaishou, Bilibili, Weibo, Tieba, and Zhihu.\\
  \quad GitHub: \url{https://github.com/NanmiCoder/MediaCrawler}
\end{itemize}

\subsection*{3) Speech and Audio Processing (ASR, VAD, Separation, End-to-End Systems)}

\begin{itemize}
  \item \textbf{speechbrain}: A PyTorch-based speech processing toolkit supporting ASR, TTS, keyword spotting, and related tasks.\\
  \quad GitHub: \url{https://github.com/speechbrain/speechbrain}

  \item \textbf{ESPnet}: An end-to-end speech processing toolkit covering ASR, TTS, and speech translation.\\
  \quad GitHub: \url{https://github.com/espnet/espnet}

  \item \textbf{silero-vad}: A pretrained voice activity detection (VAD) model optimized for real-time usage.\\
  \quad GitHub: \url{https://github.com/snakers4/silero-vad}

  \item \textbf{spleeter}: A music source separation tool for isolating vocals and accompaniment using pretrained models.\\
  \quad GitHub: \url{https://github.com/deezer/spleeter}
\end{itemize}

\subsection*{4) Vision and Video Processing and Generation}

\begin{itemize}
  \item \textbf{Segment Anything (SAM)}: A general-purpose, promptable image segmentation model.\\
  \quad GitHub: \url{https://github.com/facebookresearch/segment-anything}

  \item \textbf{Ultralytics}: Official implementations and toolchains for the YOLO family, including object detection, segmentation, and pose estimation.\\
  \quad GitHub: \url{https://github.com/ultralytics/ultralytics}

  \item \textbf{AnimeGANv3}: A framework for converting photos and videos into anime-style imagery.\\
  \quad GitHub: \url{https://github.com/TachibanaYoshino/AnimeGANv3}

  \item \textbf{transparent-background}: A background removal tool based on the InSPyReNet architecture.\\
  \quad GitHub: \url{https://github.com/plemeri/transparent-background}

  \item \textbf{Bringing Old Films Back to Life}: A system for old film restoration, published at CVPR~2022.\\
  \quad GitHub: \url{https://github.com/raywzy/Bringing-Old-Films-Back-to-Life}

  \item \textbf{PySceneDetect}: A library for video shot boundary detection and scene transition analysis.\\
  \quad GitHub: \url{https://github.com/Breakthrough/PySceneDetect}

  \item \textbf{moviepy}: A Python-based video editing framework supporting clipping, concatenation, effects, and text overlays.\\
  \quad GitHub: \url{https://github.com/Zulko/moviepy}

  \item \textbf{ffmpeg-python}: Python bindings for FFmpeg, including support for complex filter graphs.\\
  \quad GitHub: \url{https://github.com/kkroening/ffmpeg-python}

  \item \textbf{Stable Diffusion (CompVis)}: A diffusion-based text-to-image generation model.\\
  \quad GitHub: \url{https://github.com/CompVis/stable-diffusion}
\end{itemize}

\subsection*{5) Development Security and Vulnerability Detection}

\begin{itemize}
  \item \textbf{Bandit (PyCQA)}: A static analysis tool for identifying common security issues in Python codebases.\\
  \quad GitHub: \url{https://github.com/PyCQA/bandit}

  \item \textbf{trufflehog}: A tool for discovering and verifying leaked secrets in source code repositories and commit histories.\\
  \quad GitHub: \url{https://github.com/trufflesecurity/trufflehog}

  \item \textbf{sqlmap}: An automated tool for detecting and exploiting SQL injection vulnerabilities and database takeover.\\
  \quad GitHub: \url{https://github.com/sqlmapproject/sqlmap}

  \item \textbf{Bolt (s0md3v)}: A scanner designed to detect Cross-Site Request Forgery (CSRF) vulnerabilities.\\
  \quad GitHub: \url{https://github.com/s0md3v/Bolt}
\end{itemize}

\subsection*{6) NLP, String Processing, and Prompt Engineering}

\begin{itemize}
  \item \textbf{RapidFuzz}: A high-performance library for fuzzy string matching and similarity measurement.\\
  \quad GitHub: \url{https://github.com/rapidfuzz/RapidFuzz}

  \item \textbf{prompt-optimizer}: A tool for optimizing prompts to improve the quality and effectiveness of LLM interactions.\\
  \quad GitHub: \url{https://github.com/linshenkx/prompt-optimizer}
\end{itemize}

\subsection*{7) Chemistry, Molecular Analysis, and Synthesis Planning}

\begin{itemize}
  \item \textbf{AiZynthFinder}: A retrosynthetic planning system for organic chemistry based on Monte Carlo Tree Search.\\
  \quad GitHub: \url{https://github.com/MolecularAI/aizynthfinder}

  \item \textbf{ChemFormula}: A utility for parsing chemical formulas, computing molecular weights, and formatting outputs.\\
  \quad GitHub: \url{https://github.com/molshape/ChemFormula}

  \item \textbf{chemlib}: A general-purpose Python library for chemical calculations, periodic table access, and stoichiometry.\\
  \quad GitHub: \url{https://github.com/harirakul/chemlib}
\end{itemize}

\subsection*{8) Web, Backend, and General Frameworks}

\begin{itemize}
  \item \textbf{bottle}: A lightweight Python web microframework that can be deployed as a single-file application.\\
  \quad GitHub: \url{https://github.com/bottlepy/bottle}
\end{itemize}

\subsection*{9) Financial Backtesting}

\begin{itemize}
  \item \textbf{backtrader}: A quantitative trading strategy backtesting framework supporting multiple data feeds, commissions, visualization, and optimization.\\
  \quad GitHub: \url{https://github.com/mementum/backtrader}
\end{itemize}

\section{Prompts}
\label{sec:prompts}

\subsection{Agent Generation Prompt}
\label{subsec:agentization_prompt}
This section presents the prompts used to invoke four distinct frameworks for the agent generation process.

\begin{promptbox}[Claude Code]
IMPORTANT: You are working in repository directory: {cwd}

CRITICAL: All commands MUST be executed in the repository directory.
Always prefix your commands with: cd {cwd} &&

For example:
- cd {cwd} && uv sync
- cd {cwd} && source .venv/bin/activate
- cd {cwd} && python -m pytest

This ensures that virtual environments (.venv) and all generated files are created in the correct repository location.

Now, carefully read the content in {agentify_md_path} and generate an agent according to its requirements and instructions.

Specific requirements:
1. Read the agentify.md file completely and understand all phase requirements
2. Execute systematically according to Phases 1-4 in the documentation:
   - Phase 1: Environment Setup (remember to cd {cwd} before uv sync!)
   - Phase 2: Repository Analysis
   - Phase 3: Agent Setup & Customization
   - Phase 4: Testing and Deployment
3. Generate all required files:
   - skills.json (skill configuration)
   - repo_func.md (repository documentation)
   - .env (environment configuration)
   - Other necessary configuration files
4. Set default port to: {target_port}
5. Verify after completing each phase
6. Use Claude Agent SDK as the agent execution engine

Please explain each step in detail and provide test verification results after completion.
\end{promptbox}

\begin{promptbox}[Codex]
IMPORTANT: You are working in repository directory: ${this.cwd}

CRITICAL: All commands MUST be executed in the repository directory.
Always prefix your commands with: cd ${this.cwd} &&

For example:
- cd ${this.cwd} && uv sync
- cd ${this.cwd} && source .venv/bin/activate
- cd ${this.cwd} && python -m pytest

This ensures that virtual environments (.venv) and all generated files are created in the correct repository location.

Now, carefully read the content in ${this.agentifyMdPath} and generate an agent according to its requirements and instructions.

Specific requirements:
1. Read the agentify.md file completely and understand all phase requirements
2. Execute systematically according to Phases 1-4 in the documentation:
   - Phase 1: Environment Setup (remember to cd ${this.cwd} before uv sync!)
   - Phase 2: Repository Analysis
   - Phase 3: Agent Setup & Customization
   - Phase 4: Testing and Deployment
3. Generate all required files:
   - skills.json (skill configuration)
   - repo_func.md (repository documentation)
   - .env (environment configuration)
   - Other necessary configuration files
4. Set default port to: ${this.targetPort}
5. Verify after completing each phase
6. Use Codex SDK as the agent execution engine

Please explain each step in detail and provide test verification results after completion.
\end{promptbox}

\begin{promptbox}[Openhands]
IMPORTANT: First, create an isolated Python virtual environment in the current workspace using uv (if it doesn't exist):
1. Check if .venv directory already exists
2. If not, run: uv venv .venv --seed -p 3.12
3. Activate the virtual environment: source .venv/bin/activate
4. Verify Python interpreter: which python (should point to workspace's .venv/bin/python)
5. All subsequent operations MUST be executed in the activated virtual environment

Then, carefully read the requirements in {agentify_md_path}, generate a todo list. Following the todo list and requirements, automatically execute the related operations to build an a2a agent in this workspace. After building, carefully check the agent's functionality, ensure all skills are correctly loaded, and test with test_client script to ensure each skill is available. The .env template is located in templates/, default port is {port}

NOTE: All dependency installations should use 'uv pip install', and all commands must be executed in the activated virtual environment.
\end{promptbox}

\begin{promptbox}[EnvX]
You are an expert A2A agent developer. Your task is to help build an A2A-compliant agent from a code repository by following the comprehensive agentify guide.

## Working Directory: {cwd}
CRITICAL: You are working in repository directory: {cwd}
All bash commands MUST be executed in the repository directory.
Always prefix your commands with: cd {cwd} &&

Examples:
- cd {cwd} && uv sync
- cd {cwd} && source .venv/bin/activate && python -m pytest

This ensures that virtual environments (.venv) and all generated files are created in the correct repository location.

## Your Capabilities:
- Read and analyze repository code and documentation
- Create and edit Python files (agent.py, config.py, skills.json, etc.)
- Execute bash commands for testing and verification
- Search and grep code to understand structure

## Agentify Guide Location: {agentify_md_path}

## Your Task:
Carefully read the content in {agentify_md_path} and generate an agent according to its requirements and instructions.

## Specific Requirements:
1. **Read the agentify.md file completely** and understand all phase requirements
2. **Execute systematically** according to Phases 1-4 in the documentation:
   - Phase 1: Environment Setup (remember to cd {cwd} before uv sync!)
   - Phase 2: Repository Analysis
   - Phase 3: Agent Setup & Customization
   - Phase 4: Testing and Deployment

3. **Generate all required files:**
   - skills.json (skill configuration)
   - repo_func.md (repository documentation)
   - .env (environment configuration)
   - Other necessary configuration files

4. **Set default port to:** {target_port}

5. **Test as you go** - verify each phase before moving to the next

6. **Be thorough** - complete all phases from environment setup to testing

7. **Always use cd {cwd} &&** before any bash command

## Implementation Notes:
- The agent template uses Claude Agent SDK as the execution engine
- skills.json defines repository functions exposed as agent skills
- Always verify file paths and imports before creating configurations

## Your Workflow:
1. Read and understand the agentify.md guide
2. Analyze the target repository structure
3. Follow each phase in the guide systematically
4. Create all required files and configurations
5. Test the agent thoroughly

Start by carefully reading the agentify.md file, then proceed phase by phase.
Please explain each step in detail and provide test verification results after completion.
\end{promptbox}

\subsection{Agent Skill Judge Prompt}
\label{subsec:agent_skill_judge}

\paragraph{Prompt: Skill Coverage Judgement.}
This prompt judges whether a test skill is covered by a target skill set, based on actual capability.

\begin{promptbox}[Prompt: Skill Coverage Judgement]
{
  "role": "evaluator",
  "instruction": "You are a strict but pragmatic evaluator. Your task is to determine whether the Test Skill is covered (i.e., contained) by the Target Skill Set at the level of actual capability.",

  "rules": [
    "One-to-one correspondence is not required. A test skill may map to one or multiple target skills.",
    "Partial matching is allowed only if the core capability is covered.",
    "Name similarity alone is insufficient; judgement must be based on functionality, descriptions, tags, and examples.",
    "The output must be valid JSON only, with no additional text."
  ],

  "output_format": {
    "hit": "0 or 1"
  },

  "inputs": {
    "test_skill": "{json.dumps(q, ensure_ascii=False)}",
    "target_skill_set": "{json.dumps(tgt, ensure_ascii=False)}"
  }
}
\end{promptbox}

\subsection{Single-Repo Task Generation Prompts}
\label{subsec:single_repo_data_generation_prompt}
\paragraph{Prompt 1: Repository Capability Analysis and Task Ideation.}
This prompt requests a structured capability summary of the repository and a set of lightweight, file-producing evaluation tasks.

\begin{promptbox}[Prompt 1: Repository Capability Analysis and Task Ideation]
Repository location:
  <path>

Path requirement:
- Use absolute paths for every file you read or write.
- Do not use relative paths, including dot-prefixed paths.

Output file:
- Write the repository capability summary to:
  <path>

Tasks:
1) Read the repository README and determine the primary capabilities and intended usage.
2) Identify whether the repository provides a ready-to-run demo.
   - If a demo exists, document how to run it and what output artifact it produces.
3) Propose 10 simple evaluation tasks that test whether a code agent can use the repository effectively.
   - Each task must be executable end-to-end and must produce a concrete artifact file (audio/video/image/pdf/csv preferred; structured JSON is acceptable).
   - Avoid training procedures and other compute-intensive workflows.
   - Do not request a narrative "understanding" or an architecture overview as the final output.
   - Prefer tasks aligned with the repository demo(s), reusing demo inputs when applicable.
   - Specify expected inputs and a clear target artifact for each task.
   - Tasks must not require Docker.
   - Tasks must not require external APIs or secret keys.

Write the capability summary and the 10 proposed tasks to:
  <path>
\end{promptbox}

\paragraph{Prompt 2: Task Specification and Feasibility Validation.}
This prompt converts the proposed tasks into reproducible task files and validates feasibility by executing each task once.

\begin{promptbox}[Prompt 2: Task Specification and Feasibility Validation]
Path requirement:
- Use absolute paths for every file you read or write.
- Do not use relative paths, including dot-prefixed paths.

For the 10 tasks you proposed (Task 1-10), create one task specification file per task:
  <path>

For each task, you must:
1) Provide a precise task description (inputs, required operations, and expected artifact).
2) Execute the task yourself once to confirm it is feasible and can be completed successfully.
3) Define an objective success criterion based on the produced artifact file.

If a task requires input files, store them under:
  <path>

Use the following task specification format:
1. Task Description
2. Detailed Inputs
3. Your Execution Trace (not shown to the evaluated agent; write step-by-step commands and outputs)
4. Expected Output (the artifact file and key properties used for validation)

Environment requirement:
- Create and activate a new conda environment named {repo_name}.
- If conda is installed at a fixed absolute location, reference it explicitly (for example, <path>).
- Install any required tools via pip within that environment.
\end{promptbox}

\paragraph{Prompt 3: Judge Standard Extraction from Verified Runs.}
This prompt extracts concrete evaluation steps and a standardized expected response from verified execution logs.

\begin{promptbox}[Prompt 3: Judge Standard Extraction]
Path requirement:
- Use absolute paths for every file you read or write.
- Do not use relative paths, including dot-prefixed paths.

Execute tasks 1-10 again, and extract the major execution steps from the run logs using the format:
Step1: action 1, Output 1
Step2: action 2, Output 2
...

Requirements:
1) Each action must be a concrete operation (exact commands, scripts, and tools used), not a vague description.
   Save to:
     <path>
2) Also record:
   - What the final artifact looks like, and
   - What the agent should say when delivering the artifact in its final response.
   Save to:
     <path>
3) If you cannot complete a task, delete the entire task folder for that task (do not claim completion).
4) After validation, delete the {repo_name} conda environment.
\end{promptbox}

\paragraph{Prompt 4: Consolidation into a Single JSON Dataset.}
This prompt consolidates all validated tasks into a single structured JSON file suitable for downstream benchmark ingestion.

\begin{promptbox}[Prompt 4: JSON Consolidation]
Path requirement:
- Use absolute paths for every file you read or write.
- Do not use relative paths, including dot-prefixed paths.

Consolidate all tasks into a single JSON file:
  <path>

Each task entry must follow this schema:
{
  "task_id": "{repo_name}_task_{task_num}",
  "task_category": "single_agent",
  "task_description": "{the description of the task}",
  "fuzzy_description": "{fuzzy description of the task}",
  "input_parts": [
    { "kind": "text", "text": "{the description of the task}" },
    {
      "kind": "file",
      "file": {
        "path": "<path>",
        "mime_type": "..."
      }
    }
  ]
}
\end{promptbox}

\subsection{Multi-Repo Task Generation Prompts}
\label{subsec:multi_repo_data_generation_prompt}

\begin{promptbox}[Multi-Repo Task Generation Search Agent Prompt]

Your task now is: based on a set of GitHub repositories I provide, construct multiple ``multi-hop tasks.'' Specifically, for this round, generate 3 tasks, each with exactly 3 hops (3-step chains). No need to generate too many tasks.

Core requirements:

Try to ensure each hop corresponds to a different repository. Each task must require multiple steps/rounds to complete.

Every step must explicitly specify:

which repository is used,

what action is performed,

what output is produced.

The final result must be verifiable (e.g., files, logs, script return codes, data structures, etc.).

The task granularity should be as fine as possible, not vague or hand-wavy. Because the chosen repositories will likely involve many file types, an example could be: the first repository takes an input file, processes it, and passes the output artifact to later repositories. In that case, how to obtain the initial input file from the first repository becomes very important.

For this ``where do the files come from'' part, I want you to draw inspiration from each repository's README and tests. Ideally, you should directly obtain file ideas from the repositories' tests (or example assets), rather than asking me to provide missing files.

You may ask me to fetch the initial file if needed, but do not give me choices. Treat me like a robot with no autonomy; tell me exactly where to get the file. If the initial file still requires me to create or choose it myself, then forget it, that's too troublesome.

Important rules:

We may have many repositories; you do not need to parse them all at once.

You may first infer the general domain/purpose from repository names, then select the most valuable repositories that are most likely to be composable into tasks.

For the repositories you select, you may further inspect README, examples, tests, etc., to gain deeper understanding.

If information is insufficient, do not fabricate. Directly abandon that task idea instead of making things up.

Output format requirements (important)

When outputting:

You may first write 1-3 short paragraphs in natural language describing your thoughts/insights.

Then you must output a valid large JSON:

The final output must be a JSON array (list) as the top-level structure: [...].

Each element in the array is an object (dict) with the following structure:

The top level has only one key: "multi_hop_tasks".

The value of "multi_hop_tasks" is an array.

To keep granularity clear, each object contains only 1 task (i.e., the "multi_hop_tasks" array length must be 1).

For multiple tasks, add more elements to the top-level array, for example:

[
  {
    "multi_hop_tasks": [
      {
        "task_name": "...",
        "goal": "...",
        "repo_used": [
          {
            "name": "...",
            "guessed_domain": "...",
            "reason_selected": "...",
            "core_capabilities": "...",
            "i/o": "...",
            "interfaces": "CLI | API | script | demo"
          }
        ],
        "required_repos": ["repoA", "repoB", "..."],
        "multi_hop_logic": "Full chain explanation: repoA -> repoB -> repoC ...",
        "steps": [
          {
            "step": 1,
            "using_repo": "repoA",
            "action": "...",
            "expected_output": "...",
            "why_this_repo": "..."
          },
          {
            "step": 2,
            "using_repo": "repoB",
            "action": "...",
            "expected_output": "...",
            "why_this_repo": "..."
          }
        ],
        "final_artifact": "...",
        "verification": [
          "Check whether ... file exists",
          "Check whether output JSON contains field ...",
          "Check whether logs contain keyword ...",
          "Check whether script return code is 0"
        ]
      }
    ]
  },
  {
    "multi_hop_tasks": [
      { "...task 2..." : "..." }
    ]
  },
  {
    "multi_hop_tasks": [
      { "...task 3..." : "..." }
    ]
  }
]

Additional requirements:

Each task must clearly explain what the multi-hop logic is (e.g., repoA produces data -> repoB analyzes/optimizes/verifies -> repoC visualizes/deploys).

Tasks should be diverse, don't make them all the same pattern.

Do not assume repository capabilities; if uncertain, either ask questions or make conservative inferences.

When there are many repos, you may prioritize the most key/potential ones rather than processing all.

Okay-now you can start classification and processing based on the repositories below. This will be hard; you don't need to rush. You can think slowly, iterate repeatedly, and search repeatedly.

[Repositories Provided]

Document/Web parsing & OCR

Tesseract: classic open-source OCR engine in C++, multilingual, broad ecosystem. (https://github.com/tesseract-ocr/tesseract
)

EasyOCR: PyTorch OCR, out-of-the-box, 80+ languages. Good for quick integration. (https://github.com/JaidedAI/EasyOCR
)

textract: unified interface to extract text from ``almost any document'' (Word/PDF/etc., often via external engines). (https://github.com/deanmalmgren/textract
)

tika-python: Python bindings for Apache Tika, calls Tika REST to extract text/metadata. (https://github.com/chrismattmann/tika-python
)

camelot: structured table extraction from PDFs. (https://github.com/camelot-dev/camelot
)

python-docx2txt: extract text and images from .docx. (https://github.com/ankushshah89/python-docx2txt
)

Unstructured: document ETL/chunking/structuring pipelines for LLMs. (https://github.com/Unstructured-IO/unstructured
)

Web/platform crawling & text acquisition

trafilatura: extract main content and metadata from webpages; CLI/batch supported. (https://github.com/adbar/trafilatura
)

yt-dlp: multi-site audio/video downloader (YouTube etc.). (https://github.com/yt-dlp/yt-dlp
)

youtube-transcript-api: fetch YouTube captions/auto-captions without an API key. (https://github.com/jdepoix/youtube-transcript-api
)

MediaCrawler: crawler for Xiaohongshu/Douyin/Kuaishou/Bilibili/Weibo/Tieba/Zhihu, etc. (https://github.com/NanmiCoder/MediaCrawler
)

Speech/audio processing (ASR/VAD/separation/end-to-end speech)

speechbrain: PyTorch speech toolkit (ASR/TTS/KWS, etc.). (https://github.com/speechbrain/speechbrain
)

ESPnet: end-to-end speech toolkit (ASR/TTS/ST, etc.). (https://github.com/espnet/espnet
)

silero-vad: pretrained voice activity detection. (https://github.com/snakers4/silero-vad
)

spleeter: music source separation (vocals/accompaniment), pretrained. (https://github.com/deezer/spleeter
)

Vision/video processing & generation

segment-anything (SAM): general promptable segmentation model. (https://github.com/facebookresearch/segment-anything
)

Ultralytics: official YOLO toolchain for detection/segmentation/pose, etc. (https://github.com/ultralytics/ultralytics
)

AnimeGANv3: photo/video anime stylization. (https://github.com/TachibanaYoshino/AnimeGANv3
)

transparent-background: background removal matting based on InSPyReNet. (https://github.com/plemeri/transparent-background
)

Bringing-Old-Films-Back-to-Life: old film restoration (CVPR 2022). (https://github.com/raywzy/Bringing-Old-Films-Back-to-Life
)

PySceneDetect: video shot boundary / transition detection. (https://github.com/Breakthrough/PySceneDetect
)

moviepy: Python video editing (cut/concat/effects/text, etc.). (https://github.com/Zulko/moviepy
)

ffmpeg-python: Python bindings for FFmpeg (incl. complex filters). (https://github.com/kkroening/ffmpeg-python
)

Stable Diffusion (CompVis): text-to-image diffusion model. (https://github.com/CompVis/stable-diffusion
)

Dev security / vulnerability detection

Bandit: static security scanner for Python code. (https://github.com/PyCQA/bandit
)

trufflehog: find and verify leaked secrets in repos/history. (https://github.com/trufflesecurity/trufflehog
)

sqlmap: automated SQL injection testing and database takeover. (https://github.com/sqlmapproject/sqlmap
)

Bolt (s0md3v): CSRF scanner. (https://github.com/s0md3v/Bolt
)

NLP/string processing & prompt engineering

RapidFuzz: high-performance fuzzy matching/string similarity metrics. (https://github.com/rapidfuzz/RapidFuzz
)

prompt-optimizer: prompt optimizer for writing higher-quality prompts. (https://github.com/linshenkx/prompt-optimizer
)

Chemistry/molecules & synthesis planning

AiZynthFinder: MCTS-based retrosynthesis planning. (https://github.com/MolecularAI/aizynthfinder
)

ChemFormula: parse chemical formulas, compute molar mass, formatting, etc. (https://github.com/molshape/ChemFormula
)

chemlib: general chemistry computations/periodic table/stoichiometry, etc. (https://github.com/harirakul/chemlib
)

Web/backend & general frameworks

bottle: lightweight Python web micro-framework (single-file usable). (https://github.com/bottlepy/bottle
)

Financial backtesting

backtrader: quantitative strategy backtesting framework (data feeds/commission/plotting/optimization, etc.). (https://github.com/mementum/backtrader
)

\end{promptbox}

\begin{promptbox}[Multi-Repo Task Generation Code Agent Prompt]

code_agent_answer_generation_prompt = """
You are a **Code Execution Agent**. A collaborating Search/Task Agent will construct **multi-hop tasks** for you based on a set of GitHub repositories, and **your responsibility is to actually execute these tasks in the local code repositories, verify feasibility, and record execution traces**.

Your primary objective is: **to audit whether these tasks are reproducible and executable in the current environment**, rather than "forcing all steps to run no matter what."

Your core workflow is:
Read the current directory's `data.json` -> execute the tasks step by step using the code in `code_repos/` -> produce real intermediate artifacts (`artifact/`) and execution logs (`traj_log/`) -> write the final JSON (`artifact/data_execution.json`).

If, during execution, you discover that the task itself has serious issues or cannot be reasonably continued, you may **terminate early at an appropriate step**, and clearly explain the reason in the results.

## 0. Working Directory & File Layout (Hard Constraints)

You always work in a directory structured roughly as follows:

.
|-- artifact/       # Intermediate state files & final result JSON (data_execution.json)
|-- code_repos/     # All GitHub repositories live here
|   |-- code-repo-A
|   |-- code-repo-B
|   `-- code-repo-C
|-- data.json       # Multi-hop task descriptions generated by the Search/Task Agent
`-- traj_log/       # Step-level debug logs of the execution process

Conventions (VERY IMPORTANT):
- All multi-hop tasks you need to execute come from `./data.json`.
- All code must be run inside `./code_repos/<repo_name>/`.
- All intermediate and final artifacts must be written to `./artifact/`.
- All execution logs must be written to `./traj_log/` (you may choose the internal structure, but paths must be traceable in the final JSON).

## 1. Structure of data.json and Field Semantics

The overall structure of `data.json` looks roughly like this:

{
  "multi_hop_tasks": [
    {
      "task_name": "...",
      "goal": "...",

      "repo_used": [
        {
          "name": "...",
          "guessed_domain": "...",
          "reason_selected": "...",
          "core_capabilities": "...",
          "i/o": "...",
          "interfaces": "CLI | API | script | demo"
        }
      ],

      "required_repos": ["repoA", "repoB", "..."],
      "multi_hop_logic": "High-level natural language description of the multi-hop data flow: repoA -> repoB -> ...",

      "steps": [
        {
          "step": 1,
          "using_repo": "repoA",
          "action": "...",
          "expected_output": "...",
          "why_this_repo": "..."
        },
        {
          "step": 2,
          "using_repo": "repoB",
          "action": "...",
          "expected_output": "...",
          "why_this_repo": "..."
        }
      ],

      "final_artifact": "...",
      "verification": [
        "Check whether ... file exists",
        "Check whether output JSON contains field ...",
        "Check whether script exit code is 0"
      ]
    }
  ]
}

Key fields (brief explanations):
- `multi_hop_tasks`: A list where each element is an independent multi-hop task (in practice, you are usually given only one).
- `goal`: The overall intent of the task, helping you understand what the pipeline is trying to achieve.
- `repo_used`: A list describing the repositories involved, intended to give you a rough understanding of each repo's capabilities:
  - `name / guessed_domain / reason_selected / core_capabilities / i/o / interfaces`:
    These are upstream summaries and **for reference only**; the ground truth is always the actual repository files, README, and tests.
- `required_repos`: The repositories actually required for this task, typically consistent with `using_repo`.
- `multi_hop_logic`: A high-level description of the data flow (e.g., "download video -> scene detection -> editing"). After execution, you must output `multi_hop_flow` in the results; you may reference this field but do not need to copy it verbatim.
- `steps`: The ordered list of concrete steps you must execute:
  - `step`: Step index (1, 2, 3, ...).
  - `using_repo`: The repository to use for this step, corresponding to `./code_repos/<using_repo>/`.
  - `action`: A human-language instruction that you must translate into concrete executable commands/scripts.
  - `expected_output`: The ideal output, used to judge success.
  - `why_this_repo`: Why this repository is used, to clarify design intent.
- `final_artifact`: A natural language description of the ideal final output. You should aim to approximate this and record the actual artifact path in your results.
- `verification`: A task-level verification checklist. In `data_execution.json`, you must mark each item as pass / fail / not_applicable, with brief evidence.

## 2. Environment & Tooling Conventions (uv)

Before being provided to you, all repositories **are typically already configured with uv environments** in their respective directories, in the form:

- `./code_repos/<repo_name>/.venv/`

Whenever you need to execute commands using uv inside a repository, you **must follow this fixed pattern** (RepoA as an example):

- After entering the repository directory, the actual command should look like:

  ```bash
  cd ./code_repos/RepoA
  source .venv/bin/activate && uv run python your_script.py ...
That is: before every uv run, you must first source .venv/bin/activate, and these actions must be bound to the same command or the same shell session.

If, even after activating .venv, uv run cannot execute properly (e.g., missing dependencies), you may perform the minimum necessary uv installation steps based on repository files (pyproject.toml, requirements, setup, etc.).
Do not over-install or rebuild environments unless necessary.

Regardless of whether dependencies are added, you must record all environment-related commands you executed (if any) in both logs and the final JSON.

3. Execution Principles (Core Responsibilities & Workflow)
3.0 Overall Workflow
Your execution logic is straightforward:
Read data.json -> execute steps in order in the corresponding repositories -> write artifacts to artifact/ -> write logs to traj_log/ -> judge task feasibility -> write data_execution.json.

You are not responsible for redesigning the task. Your job is to run the upstream-defined steps as faithfully as possible and leave a clear execution trail.

If, during execution, you determine that the task design, local repositories, or dependencies have serious issues such that continuing is meaningless or too risky, you may stop further execution and explain your reasoning directly in the results.

3.1 Sequential Execution & Real Artifacts
Steps must be executed sequentially (unless you have already decided to terminate early).

All files produced at each step (text, CSV, JSON, videos, etc.) must be written to artifact/ and recorded with paths in the result JSON.

Subsequent steps must explicitly consume actual outputs from previous steps, not fabricated objects.

In data_execution.json, summarize the actual multi-hop data flow (multi_hop_flow) in natural language, including which step execution reached and where it stopped (if applicable).

3.2 Logging (Reproducibility-Oriented)
Write stdout/stderr of all commands to traj_log/<task_name>/.

For each command, record: cmd, cwd, exit_code, and the two log file paths.

Logs do not need to be pretty; they only need to be reproducible.

3.3 No Fabrication (Strictly Based on Actual Repo Structure)
Only use scripts, data, and examples that actually exist in the repositories (README / examples / tests).

If a file mentioned in the task description cannot be found in the repository:

Record what is missing and where you searched,

Then terminate execution, as this indicates a serious flaw in the task design.

3.4 Error Handling (Explainability > Forced Success)
When a command fails, clearly record: command, cwd, exit_code, and log paths.

You may make small, reasonable, and explainable fixes (e.g., adjusting parameters or paths), but must record the reason.

If, after multiple attempts, you conclude that the task design is flawed, dependencies are fundamentally unsatisfiable, or continuing execution is meaningless, you may:

Stop executing further steps;

Mark the task status as partial_success or failed;

Clearly explain why you stopped, which step you reached, and what upstream needs to fix or clarify in multi_hop_flow, overall_verification, and/or clarification_requests_if_any.

4. Output Format (Returned JSON & Persistence)
After processing all multi_hop_tasks (including failure or early termination cases), you must:

Generate a JSON file:

Path: ./artifact/data_execution.json

Content: execution-related information only; do not repeat design-time fields already present in data.json (e.g., original step text, required_repos, why_this_repo).

At the same time, return the full contents of data_execution.json as your response to the upstream agent.

Requirements for data_execution.json:

Do not include design fields already present in data.json;

Retain only: execution traces, artifact paths, log paths, verification results, and your subjective judgment and feedback on task validity.

You must explicitly add two fields to express your overall judgment of the task:

task_verification (boolean):

true: In your judgment, the task is basically reproducible/executable in the current environment (minor non-critical failures allowed).

false: You believe the task has serious issues (e.g., missing key files, unavailable core dependencies, self-contradictory design) and cannot be considered valid.

task_verification_summary (string):

If true: briefly explain how the task was completed and whether key verification items passed.

If false: state which step failed, what issue occurred, and why the task design or dependencies are problematic.

An example structure (you may extend it, but do not remove these core fields):

json
Copy code
{
  "tasks": [
    {
      "task_verification": true,
      "task_verification_summary": "Successfully executed through step 3; verification checklist items on file existence and return codes passed. The task is reproducible in the current environment.",

      "task_name": "...",
      "goal": "...",
      "status": "success | partial_success | failed",

      "multi_hop_flow": "Natural language explanation of the actual multi-hop data flow (including how far execution reached and where it stopped, with reasons)",

      "steps": [
        {
          "step": 1,
          "repo": "RepoA",
          "repo_path": "./code_repos/RepoA",

          "interpretation": "How you translated the natural language action into concrete commands and which actual files were used",

          "commands": [
            {
              "cmd": "...",
              "cwd": "...",
              "exit_code": 0,
              "stdout_log": "traj_log/<task_name>/step1_cmd1_out.log",
              "stderr_log": "traj_log/<task_name>/step1_cmd1_err.log"
            }
          ],

          "artifacts": [
            {
              "path": "artifact/step1_output.txt",
              "description": "Artifacts produced in this step that are used by subsequent steps (even if execution stops early, record what was produced)"
            }
          ],

          "verification": [
            {
              "item": "Check whether artifact/step1_output.txt exists",
              "result": "pass | fail",
              "evidence": "E.g., file size or brief ls output"
            }
          ]
        }
      ],

      "final_artifact": {
        "path": "artifact/<task_name>_final_output.xxx",
        "description": "What you consider the most critical final result file (or note that no final artifact was produced if the task was not completed)"
      },

      "overall_verification": {
        "summary": "Whether the task satisfies the verification conditions defined in data.json (success / partial / failure, with reasons)",
        "items": [
          {
            "item": "Check whether JSON contains field ...",
            "result": "pass | fail | not_applicable",
            "evidence": "Brief structural or content summary"
          }
        ]
      },

      "logs_root": "traj_log/<task_name>/",
      "artifacts_root": "artifact/"
    }
  ],

  "clarification_requests_if_any": [
    "If you believe the task design, local repositories, or dependencies are problematic and prevent proper execution, list here-in natural language-what you want the upstream agent to fix or clarify."
  ]
}
5. Style Preferences (Additional Notes)
In the interpretation field, explain how you concretely translated natural language actions into commands, avoiding "black-box" execution.

Prefer using repository-provided tests/, examples/, or demo/ inputs and command patterns, rather than inventing inputs from scratch.

Respect task diversity and upstream design choices; do not proactively simplify all tasks into "just run a single script," as long as the task goal remains unchanged.

Do not fabricate success just to mark status as success: as long as your execution trace is real and reproducible, even a failed or early-terminated trajectory is valuable audit data.
"""
\end{promptbox}

\section{Human Check Criteria on Tasks}
This section presents the human check criteria on single-repo tasks and multi-repo tasks.

\subsection{Single-Repo Tasks Human Check Criteria}
\label{subsec:single_repo_tasks_check_criteria}
\begin{promptbox}[Single-Repo Tasks Human Check Criteria]
Core Capability Inheritance

A valid skill must reflect a principal and non-trivial capability of the repository. It should inherit the repository's core purpose rather than surface-level or incidental behaviors.

Included examples:

a web automation repository: ``navigate webpages and extract structured content from dynamic interfaces'';
a code analysis tool: ``analyze repository dependency graphs and identify impacted modules.''

Excluded examples:

``read files'';
``parse JSON'';
``send HTTP requests'';
``log execution results.''

These excluded examples are implementation utilities rather than repository-defining capabilities.

Atomic Functional Unit

Each skill should be expressed at the granularity of one standalone functional unit. Annotators must avoid both over-broad and over-fragmented labels.

Too broad:

``perform end-to-end software engineering tasks.''

Too narrow:

``open a file,'' ``split a string,'' ``call a helper function.''

Preferred:

``generate unit tests for a given code module'';
``retrieve relevant API documentation for a user-specified library'';
``execute browser interactions grounded in webpage state.''

When a repository supports a pipeline with multiple tightly coupled stages, annotators should determine whether these stages form one indivisible functional unit or multiple reusable skills. The default preference is to split only when each part has independent semantic meaning and could plausibly be reused in isolation.

Semantic Meaningfulness

A task is semantically meaningful if it defines a coherent and understandable objective that a competent evaluator could interpret consistently. In particular, a meaningful task should satisfy the following:

the instruction is understandable without requiring hidden assumptions;
the task objective is non-empty and non-degenerate;
the requested outcome is plausible in the context of the repository and environment;
the task is neither purely decorative nor artificially constructed without practical functional content.

Tasks should be rejected under this criterion if they are:

nonsensical or self-contradictory;
too vague to determine what success would mean;
trivial to the point of not evaluating any substantive capability;
unnatural artifacts of automatic generation rather than plausible usage scenarios

Examples of semantic issues include:

the task references entities or files that do not exist;
the instruction bundles incompatible goals into one prompt;
the task is phrased so vaguely that multiple incompatible interpretations are equally plausible;
the task asks for an outcome with no discernible utility.
 Strict Dependence on Repository-Specific Capabilities

A task strictly depends on repository-specific capabilities if solving it genuinely requires the distinctive functions implemented by the target repository, rather than generic reasoning, basic scripting, or broadly available utilities.

A task should be retained only when the repository is materially necessary for successful completion. Reviewers should ask:

Would this task still be solvable in approximately the same way without this repository?
Does the task exercise a core capability represented in the repository's gold-standard skills?
Is the repository's contribution central to task completion, rather than incidental?

Tasks should be rejected if:

the task can be completed through generic language-model reasoning alone;
the task only requires common utilities such as file reading, simple string processing, or basic shell commands;
the repository could be replaced by many unrelated tools without changing the essence of the task;
the repository is only weakly relevant, while the main challenge lies elsewhere.

The key requirement is necessity, not mere relevance. It is not enough that the repository could be helpful; the task must be designed so that the repository's specific capabilities are meaningfully required.
\end{promptbox}

\subsection{Multi-Repo Tasks Human Check Criteria}
\label{subsec:multi_repo_tasks_check_criteria}
\begin{promptbox}[Multi-Repo Tasks Human Check Criteria]
Repository Complementarity

A valid multi-repository task should involve repositories that fulfill distinct and complementary roles within the overall workflow. The participating repositories should represent different functional capabilities that combine to achieve an outcome that no single repository could accomplish alone.

The relationship between repositories should be one of meaningful collaboration rather than redundancy. Each repository should contribute a unique capability or perspective that is not substantially duplicated by other participants in the task.

Tasks should generally be screened out under this criterion if the repositories involved perform largely overlapping functions, if any repository's contribution is marginal or incidental to the overall objective, or if the task could be reasonably reframed to use fewer repositories without substantial loss of functionality.


Data Flow Continuity and Verifiability

A multi-repository task should exhibit a coherent and traceable flow of data or state transformations across its constituent steps. Each step in the task should accept inputs that are meaningfully connected to the outputs of preceding steps, and the overall progression should reflect a logical sequence of operations.

The intermediate and final outputs of the task should be verifiable in a manner that allows for reasonable assessment of whether the task objectives have been met. The relationship between what a step produces and what subsequent steps consume should be discernible from the task design.

Tasks should generally be screened out under this criterion if the data flow between steps lacks clear logical connection, if steps operate in isolation without meaningful coordination, or if the relationship between step outputs and task objectives is difficult to establish.


Step Necessity

Every repository and its corresponding step in a multi-repository task should serve an essential function within the overall pipeline. Each step should represent a substantive contribution that advances the task toward its final objective.

The importance of each step should be evident from the task design. No step should appear optional, trivial, or easily bypassable without meaningful impact on the task outcome.

Tasks should generally be screened out under this criterion if any step appears to be included artificially without clear purpose, if a step's contribution is disproportionately small relative to its inclusion, or if the task could achieve substantially the same outcome by omitting or skipping a particular step.


\end{promptbox}

\end{document}